\documentclass[11pt,a4paper]{article}
\usepackage{epsfig}

 \newcommand{\be}{\begin{equation}}
 \newcommand{\ee}{\end{equation}}
 \newcommand{\ba}{\begin{eqnarray}}
 \newcommand{\ea}{\end{eqnarray}}
 \newcommand{\bl}{\begin{equation}\begin{array}{ll}}
 \newcommand{\el}{\end{array}\end{equation}}
 \newcommand{\bll}{\begin{equation}\begin{array}{lll}}
 \newcommand{\bdm}{\begin{displaymath}}
 \newcommand{\edm}{\end{displaymath}}
 \def\bea{\begin{eqnarray}}
 \def\eea{\end{eqnarray}}
 \def\barr{\begin{array}}
 \def\earr{\end{array}}

\def\lim{\rightarrow}

\begin{document}
\raggedbottom

\title{{\bf The vicinity of the phase transition in the  lattice Weinberg -
Salam Model }}

\author{M.A.~Zubkov \thanks{zubkov@itep.ru}\\{\small \it $^*$ ITEP, B.Cheremushkinskaya 25, Moscow, 117259, Russia}}

\maketitle

\begin{abstract}

We investigated the lattice Weinberg - Salam model without fermions for the
Higgs mass around $300$ GeV.  On the phase diagram there exists the vicinity of
the phase transition between the physical Higgs phase and the unphysical
symmetric phase, where the fluctuations of the scalar field become strong while
Nambu monopoles are dense. According to our numerical results (obtained on the
lattices of sizes up to $20^3\times 24$) the maximal value of the ultraviolet
cutoff in the model cannot exceed the value around $1.4$ TeV.

\end{abstract}

\section{Introduction}

Nambu monopoles are not described by means of a perturbation expansion around
the trivial vacuum background. Therefore, nonperturbative methods should be
used in order to investigate their physics. However, their mass is estimated at
the Tev scale. That's why at the energies much less than $1$ Tev their effect
on physical observables is negligible. However, when energy of the processes
approaches $1$ Tev we expect these objects influence the dynamics.

The phase diagram of  lattice Weinberg - Salam model contains physical Higgs
phase, where scalar field is condensed and gauge bosons $Z$ and $W$ acquire
their masses. This physical phase is bounded by the phase transition surface.
Crossing this surface one leaves the physical phase and enters the phase of the
lattice theory that has nothing to do with the conventional continuum
Electroweak theory. In the physical phase of the theory the Electroweak
symmetry is broken spontaneously while in the unphysical phase the  Electroweak
symmetry is not broken.

Moving along the line of constant physics in the direction of the increase of
the ultraviolet cutoff $\Lambda = \pi/a$ ($a$ is the lattice spacing) we
approach the transition surface\footnote{The line of constant physics on the
phase diagram of lattice model is the line along which physical observables are
constant while the ultraviolet cutoff is not.}. We find the indications that
there exists the maximal possible ultraviolet cutoff $\Lambda_c$ within the
physical phase. Our estimate (for the Higgs mass $M_H \sim 300$ Gev) is
$\Lambda_c \sim
 1.4$ TeV. It is important to compare this result with the limitations on the Ultraviolet
Cutoff, that come from the perturbation theory.  The latter appear as a
consequence of the triviality problem, which is related to Landau pole in
scalar field self coupling $\lambda$. Due to the Landau pole the renormalized
$\lambda$ is zero, and the only way to keep it equal to its measured value is
to impose the limitation on the cutoff. That's why the Electroweak theory is
usually thought of as a finite cutoff theory. For small Higgs masses (less than
about $350$ Gev) the correspondent energy scale calculated within the
perturbation theory is much larger, than $1$ Tev. In particular, for $M_H \sim
300$ Gev this value is about $10^3$ TeV.

On the tree level the W-boson (Z- boson) mass in lattice units vanishes on the
transition line at small enough $\lambda$. This means that the tree level
estimate predicts the appearance of an infinite ultraviolet cutoff at the
transition point for small $\lambda$. At infinite $\lambda$ the tree level
estimate gives nonzero values of lattice $M_W, M_Z$ at the transition point.
Our numerical investigation of $SU(2)\otimes U(1)$ model (at $\lambda = 0.009$)
and previous calculations in the $SU(2)$ Gauge Higgs model (both at finite
$\lambda$ and at $\lambda = \infty$) show that for the considered lattice sizes
renormalized masses do not vanish and the transition is either of the first
order or a crossover. (Actually, the situation, when the cutoff tends to
infinity at the position of the transition point means that there is a second
order phase transition.)

In table 1 of \cite{BVZ2007} the data on the ultraviolet cutoff achieved in
selected lattice studies of the $SU(2)$ Gauge Higgs model are presented.
Everywhere $\beta$ is around $\beta \sim 8$ and the renormalized fine structure
constant is around $\alpha \sim 1/110$. This table shows that the maximal value
of the cutoff ${\Lambda} = \frac{\pi}{a}$ ever achieved in these studies is
around $1.4$ Tev.

Possible explanation of the mentioned discrepancy between lattice results and
the results given by the perturbation theory is that in some vicinity of the
transition the perturbation theory does not work. Indeed we find that there
exists the vicinity of the phase transition between the Higgs phase and the
symmetric phase in the Weinberg - Salam model, where the fluctuations of the
scalar field become strong and the perturbation expansion around trivial vacuum
cannot be applied. As it was mentioned above, the continuum theory is to be
approached within the vicinity of the phase transition, i.e. the cutoff is
increased along the line of constant physics when one approaches the point of
the transition. That's why the conventional prediction on the value of the
cutoff admitted in the Standard Model based on the perturbation theory may be
incorrect.

The nature of the fluctuational region is illustrated by the behavior of
quantum Nambu monopoles \cite{Nambu,Chernodub_Nambu}. We show that their
lattice density increases when the phase transition point is approached. Within
the FR these objects are so dense that it is not possible at all to speak of
them as of single monopoles. Namely, within this region the average distance
between the Nambu monopoles is of the order of their size. Such complicated
configurations obviously have nothing to do with the conventional vacuum used
in the continuum perturbation theory.

\section{The lattice model under investigation}
The lattice Weinberg - Salam Model without fermions contains  gauge field
${\cal U} = (U, \theta)$ (where $ \quad U
 \in SU(2), \quad e^{i\theta} \in U(1)$ are
realized as link variables), and the scalar doublet $ \Phi_{\alpha}, \;(\alpha
= 1,2)$ defined on sites.

The  action is taken in the form
\begin{eqnarray}
 S & = & \beta \!\! \sum_{\rm plaquettes}\!\!
 ((1-\mbox{${\small \frac{1}{2}}$} \, {\rm Tr}\, U_p )
 + \frac{1}{{\rm tg}^2 \theta_W} (1-\cos \theta_p))+\nonumber\\
 && - \gamma \sum_{xy} Re(\Phi^+U_{xy} e^{i\theta_{xy}}\Phi) + \sum_x (|\Phi_x|^2 +
 \lambda(|\Phi_x|^2-1)^2), \label{S}
\end{eqnarray}
where the plaquette variables are defined as $U_p = U_{xy} U_{yz} U_{wz}^*
U_{xw}^*$, and $\theta_p = \theta_{xy} + \theta_{yz} - \theta_{wz} -
\theta_{xw}$ for the plaquette composed of the vertices $x,y,z,w$. Here
$\lambda$ is the scalar self coupling, and $\gamma = 2\kappa$, where $\kappa$
corresponds to the constant used in the investigations of the $SU(2)$ gauge
Higgs model. $\theta_W$ is the Weinberg angle. Bare fine structure
 constant $\alpha$ is expressed through $\beta$ and $\theta_W$ as $\alpha = \frac{{\rm tg}^2 \theta_W}{\pi \beta(1+{\rm tg}^2
\theta_W)}$. We consider the region of the phase diagram with  $\beta \sim 12$
and $\theta_W \sim \pi/6$. Therefore, bare couplings are ${\rm sin}^2 \theta_W
\sim 0.25$; $\alpha \sim \frac{1}{150}$. These values are to be compared with
the experimental ones ${\rm sin}^2 \theta_W(100 {\rm Gev}) \sim 0.23$;
$\alpha(100 {\rm Gev}) \sim \frac{1}{128}$. All simulations were performed on
lattices of sizes $8^3\times 16$, $12^3\times 16$, and $16^4$. The transition
point was checked using the larger lattice ($20^3\times 24$).

\section{Nambu monopoles}

Nambu monopoles are defined as the endpoints of the $Z$-string \cite{Nambu}.
The $Z$-string is the classical field configuration that represents the object,
which is characterized by the magnetic flux extracted from the $Z$-boson field.
The size of Nambu monopoles was estimated \cite{Nambu} to be of the order of
the inverse Higgs mass, while its mass should be of the order of a few TeV.
According to \cite{Nambu} Nambu monopoles may appear only in the form of a
bound state of a monopole-antimonopole pair. In lattice theory the following
variables are considered as creating the $Z$ boson: $ Z_{xy} = Z^{\mu}_{x} \;
 = - {\rm sin} \,[{\rm Arg} (\Phi_x^+U_{xy} e^{i\theta_{xy}}\Phi_y) ],$
and: $ Z^{\prime}_{xy} = Z^{\mu}_{x} \;
 = - \,[{\rm Arg} (\Phi_x^+U_{xy} e^{i\theta_{xy}}\Phi_y) ]$.

The classical solution corresponding to a $Z$-string should be formed around
the $2$-dimensional topological defect which is represented by the
integer-valued field defined on the dual lattice $ \Sigma = \frac{1}{2\pi}^*([d
Z^{\prime}]_{{\rm mod} 2\pi} - d Z^{\prime})$. (Here we used the notations of
differential forms on the lattice. For a definition of those notations see, for
example, ~\cite{forms}.) Therefore, $\Sigma$ can be treated as the worldsheet
of a {\it quantum} $Z$-string\cite{Chernodub_Nambu}. Then, the worldlines of
quantum Nambu monopoles appear as the boundary of the $Z$-string worldsheet: $
j_Z = \delta \Sigma $.

For historical reasons in lattice simulations we fix unitary gauge $\Phi_2 =
0$; $\Phi_1 \in {\cal R}$; $\Phi_1 \ge 0$ (instead of the usual $\Phi_1 = 0$;
$\Phi_2 \in {\cal R}$), and the lattice Electroweak theory becomes a lattice
$U(1)$ gauge theory with the $U(1)$ gauge field $A_{xy}  =  A^{\mu}_{x} \;
 = \,[Z^{\prime}  + 2\theta_{xy}]  \,{\rm mod}
 \,2\pi, $
(The usual lattice Electromagnetic field is related to $A$ as $ A_{\rm EM}  = A
- Z^{\prime} + 2 \,{\rm sin}^2\, \theta_W Z^{\prime}$.) One may try to extract
monopole trajectories directly from $A$. The monopole current is given by
%
%
$
 j_{A} = \frac{1}{2\pi} {}^*d([d A]{\rm mod}2\pi).
$
%
Both $j_Z$, and $j_A$ carry magnetic charges. That's why it is important to
find the correspondence between them.
In continuum notations we have
$
 A^{\mu}  =  Z^{\mu} + 2 B^{\mu},
$
where $B$ is the hypercharge field. Its strength is divergenceless. As a result
in continuum theory the net $Z$ flux emanating from the center
 of the monopole is equal to the net $A$ flux.
(Both $A$ and $Z$ are undefined inside the monopole.)  This means that in the
continuum limit the position of the Nambu monopole must coincide
 with the position of the monopole extracted from the field $A$.
Therefore, one can consider $j_A$  as another definition of a quantum Nambu
monopole \cite{VZ2008}. Actually, in our numerical simulations we use this
definition.

\section{Phase diagram}

\begin{figure}
\begin{center}
 \epsfig{figure=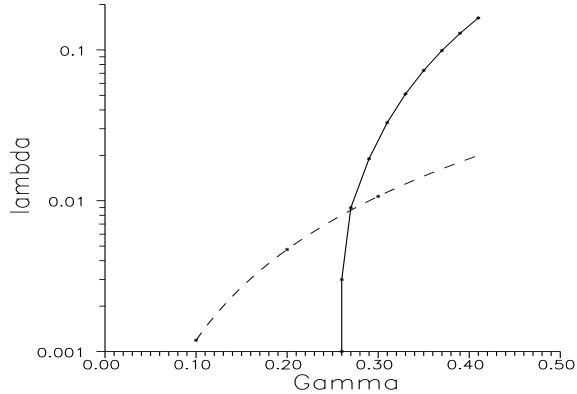,height=60mm,width=80mm,angle=0}
\caption{\label{fig.2} The phase diagram of the model in the
 $(\gamma, \lambda)$-plane at $\beta = 12$. The dashed line is the tree - level
  estimate for the line of constant physics correspondent to bare $M^0_H = 270$ Gev. The continuous line
  is the line of phase transition between the physical Higgs phase and the unphysical symmetric phase (statistical errors for
  the values of $\gamma$ on this line are about $0.005$).   }
\end{center}
\end{figure}

 In the three - dimensional ($\beta, \gamma, \lambda$) phase
diagram the transition surfaces are two - dimensional. The lines of constant
physics on the tree level are the lines ($\frac{\lambda}{\gamma^2} = \frac{1}{8
\beta} \frac{M^2_H}{M^2_W} = {\rm const}$; $\beta = \frac{1}{4\pi \alpha}={\rm
const}$). We suppose that in the vicinity of the transition  the deviation of
the lines of constant physics from the tree level estimate may be significant.
However,  qualitatively their behavior is the same. Namely, the cutoff is
increased along the line of constant physics when $\gamma$ is decreased and the
maximal value of the cutoff is achieved at the transition point. Nambu monopole
density in lattice units is also increased when the ultraviolet cutoff is
increased.

At $\beta = 12$ (corresponds to bare $\alpha \sim 1/150$) the phase diagram is
represented on Fig. \ref{fig.2}. The physical Higgs phase is situated right to
the transition line. The position of the transition is localized at the point
where the susceptibility extracted from the Higgs field creation operator
achieves its maximum.
 We use the
susceptibility  $\chi = \langle H^2 \rangle - \langle H\rangle^2$ extracted
from $H = \sum_{y} Z^2_{xy}$ (see, for example, Fig. \ref{fig.6_}). We observe
no difference between the values of the susceptibility calculated using the
lattices of the sizes $8^3\times 16$, $12^3\times 16$, and $16^4$. This
indicates that the transition may be a crossover.

It is worth mentioning that the value of the renormalized Higgs boson mass does
not deviate significantly from its bare value. For example, for $\lambda$
around $0.009$ and $\gamma$ in the vicinity of the phase transition  bare value
of the Higgs mass is around $270$ Gev while the observed renormalized value is
 $300 \pm 70$ Gev.

\section{Effective constraint potential}

\begin{figure}
\begin{center}
 \epsfig{figure=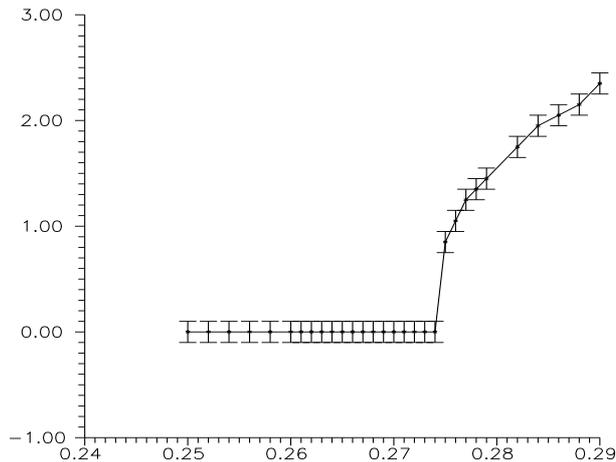,height=60mm,width=80mm,angle=0}
\caption{\label{fig.4_} $\phi_m$  as a function of $\gamma$ at $\lambda =0.009$
and $\beta = 12$.  }
\end{center}
\end{figure}

We have calculated the constraint effective potential for $|\Phi|$ using the
histogram method. The calculations have been performed on the lattice
$8^3\times 16$. The probability $h(\phi)$ to find the value of $|\Phi|$ within
the interval $[\phi-0.05;\phi+0.05)$ has been calculated for $\phi = 0.05 +
N*0.1$, $N = 0,1,2, ...$ This probability is related to the effective potential
as $ h(\phi) = \phi^3 e^{-V(\phi)}$. That's why we extract the potential from
$h(\phi)$ as
\begin{equation}
V(\phi) = - {\rm log}\, h(\phi) + 3 \, {\rm log} \, \phi \label{CEP}
\end{equation}
Next, we introduce the useful quantity $H = V(0) - V(\phi_m)$, which is called
the potential barrier hight (here $\phi_m$ is the point, where $V$ achieves its
minimum).

\begin{figure}
\begin{center}
 \epsfig{figure=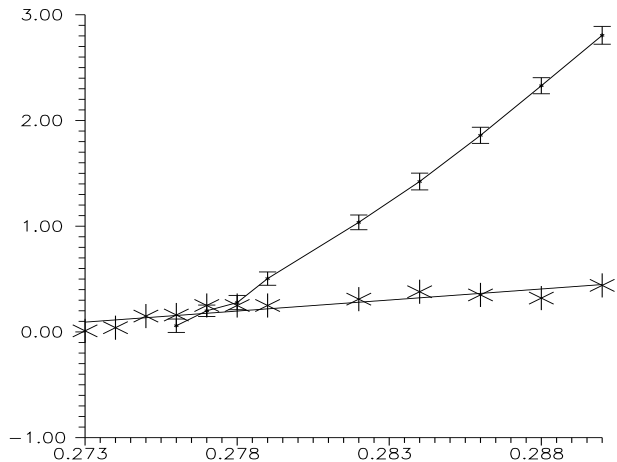,height=60mm,width=80mm,angle=0}
\caption{\label{fig.3_} $H$ (points) vs. $H_{fluct}$ (stars)  as a function of
$\gamma$ at $\lambda =0.009$ and $\beta = 12$.  }
\end{center}
\end{figure}

\begin{figure}
\begin{center}
 \epsfig{figure=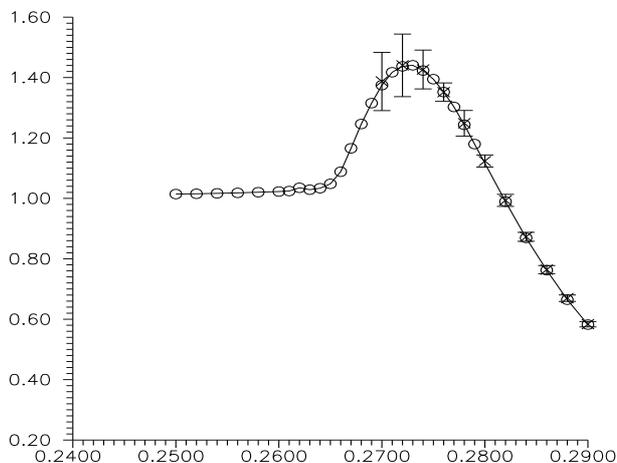,height=60mm,width=80mm,angle=0}
\caption{\label{fig.6_} Susceptibility $\langle H^2 \rangle - \langle
H\rangle^2$ (for $H_x = \sum_{y} Z^2_{xy}$) as a function of $\gamma$ at
$\lambda =0.009$ and $\beta = 12$. Circles correspond to the lattice
$8^3\times16$. Crosses correspond to the lattice $12^3\times 16$. }
\end{center}
\end{figure}

On Fig. \ref{fig.4_}  we represent the values of $\phi_m$ for $\lambda =
0.009$, $\beta = 12$.  On Fig. \ref{fig.3_} we represent the values of $H$ for
$\lambda = 0.009$, $\beta = 12$. One can see that the values of $\phi_m$ and
$H$ increase when $\gamma$ is increased. At $\gamma = 0.274$, $\lambda = 0.009$
the minimum of the potential is at $\phi = 0$. This point corresponds to the
maximum of the susceptibility constructed of the Higgs field creation operator
(see Fig. \ref{fig.6_}).
 At $\gamma = 0.275$, $\lambda = 0.009$ minimum
 of the potential is observed at nonzero $\phi_m$. That's why we localize the
position of the transition point at $\gamma = 0.273\pm 0.002$.

It is important to understand which value of barrier hight can be considered as
small and which value can be considered as large. Our suggestion is to compare
$H = V(0) - V(\phi_m)$ with $H_{\rm fluct} = V(\phi_m + \delta \phi) -
V(\phi_m)$, where $\delta \phi$ is the fluctuation of $|\Phi|$. From Fig.
\ref{fig.3_} it is clear that there exists the value of $\gamma$ (we denote it
$\gamma_{c2}$) such that at $\gamma_c < \gamma < \gamma_{c2}$ the barrier hight
$H$ is of the order of $H_{\rm fluct}$ while for $\gamma_{c2} << \gamma$ the
barrier hight is essentially larger than $H_{\rm fluct}$. The rough estimate
for this pseudocritical value is $\gamma_{c2} \sim 0.278$. We estimate the
fluctuations of $|\Phi|$ to be around $\delta \phi \sim 0.6$ for all considered
values of $\gamma$ at $\lambda = 0.009$, $\beta = 12$. It follows from our data
that $\phi_m, \langle |\phi|\rangle
>> \delta \phi$ at $\gamma_{c2} << \gamma$ while $\phi_m \sim \delta \phi$ at
$\gamma_{c2} > \gamma$. Basing on these observations we expect that in the
region $\gamma_{c2} << \gamma$ the usual perturbation expansion around trivial
vacuum of spontaneously broken theory can be applied to the lattice Weinberg -
Salam model while in the FR $\gamma_c < \gamma < \gamma_{c2}$ it cannot be
applied. At the value of $\gamma$ equal to $\gamma_{c2}$ the calculated value
of the cutoff is $1.0 \pm 0.1 $ Tev.

\section{The renormalized coupling}

\begin{figure}
\begin{center}
 \epsfig{figure=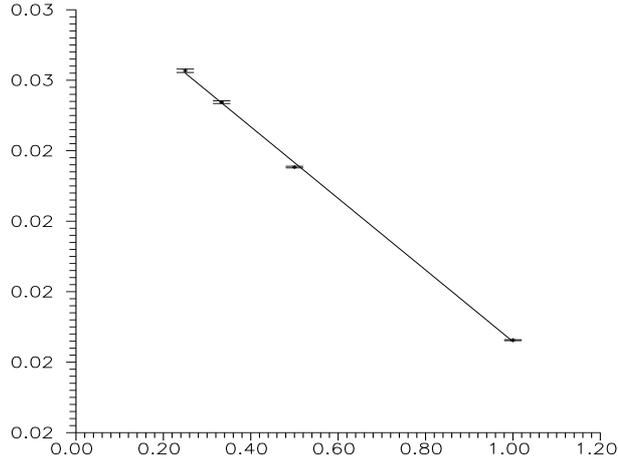,height=60mm,width=80mm,angle=0}
\caption{\label{fig.1_1_} The potential for the right - handed leptons vs.
$1/R$ at $\gamma = 0.277$, $\lambda =0.009$, and $\beta = 12$ (lattice
$8^3\times 16$). }
\end{center}
\end{figure}

In order to calculate the renormalized fine structure constant $\alpha_R =
e^2/4\pi$ (where $e$ is the electric charge) we use the potential for
infinitely heavy external fermions. We consider Wilson loops for the
right-handed external leptons:
\begin{equation}
 {\cal W}^{\rm R}_{\rm lept}(l)  =
 \langle {\rm Re} \,\Pi_{(xy) \in l} e^{2i\theta_{xy}}\rangle.
\label{WR}
\end{equation}
Here $l$ denotes a closed contour on the lattice. We consider the following
quantity constructed from the rectangular Wilson loop of size $r\times t$:
\begin{equation}
 {\cal V}(r) = {\rm log}\, {\rm lim}_{t \rightarrow \infty}
 \frac{  {\cal W}(r\times t)}{{\cal W}(r\times (t+1))}.
\end{equation}
Due to exchange by virtual photons at large enough distances we expect the
appearance of the Coulomb interaction
\begin{equation}
 {\cal V}(r) = -\frac{\alpha_R}{r} + const. \label{V1}
\end{equation}
On Fig. \ref{fig.1_1_} we represent as an example the dependence of the
potential on $1/R$ for $\gamma = 0.277$, $\lambda =0.009$, and $\beta = 12$. In
the vicinity of the transition the fit (\ref{V1}) gives values of renormalized
fine structure constant around $1/100$. The calculated values are to be
compared with bare constant $\alpha_0 = 1/(4\pi \beta)\sim 1/150$ at $\beta =
12$. However, for $\gamma >> \gamma_{c2}$ the tree level estimate is
approached. This is in correspondence with our supposition that the
perturbation theory cannot be valid within the FR while it works well far from
the FR.

\section{Masses and the lattice spacing}
\begin{figure}
\begin{center}
 \epsfig{figure=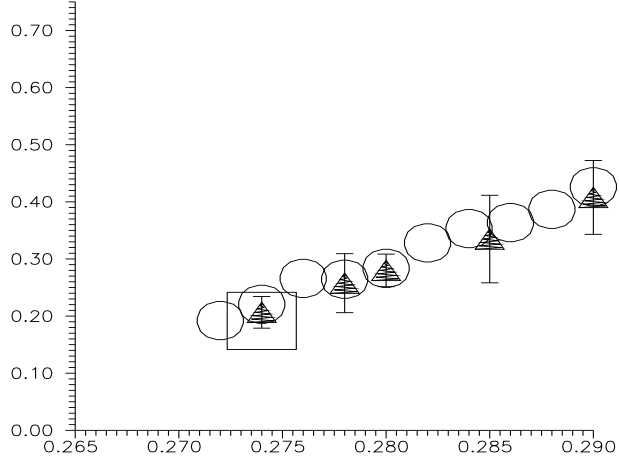,height=60mm,width=80mm,angle=0}
\caption{\label{fig.3} Z - boson mass in lattice units at $\lambda =0.009$ and
$\beta = 12$ as a function of $\gamma$. Circles correspond to lattice
$12^3\times 16$. Triangles correspond to lattice $16^4$. Squares correspond to
lattice $20^3\times 24$ (the error bars are about of the same size as the
symbols used).}
\end{center}
\end{figure}

From the very beginning we fix the unitary gauge $\Phi_1 = const.$, $\Phi_2 =
0$. The following variables are considered as creating a $Z$ boson and a $W$
boson, respectively:
\begin{eqnarray}
  Z_{xy} & = & Z^{\mu}_{x} \;
 = -{\rm sin} \,[{\rm Arg} U_{xy}^{11} + \theta_{xy}],
\nonumber\\
 W_{xy} & = & W^{\mu}_{x} \,= \,U_{xy}^{12} e^{-i\theta_{xy}}.\label{Z1}
\end{eqnarray}
Here, $\mu$ represents the direction $(xy)$.

After fixing the unitary gauge the electromagnetic $U(1)$ symmetry remains:
\begin{eqnarray}
 U_{xy} & \rightarrow & g^\dag_x U_{xy} g_y, \nonumber\\
 \theta_{xy} & \rightarrow & \theta_{xy} -  \alpha_y/2 + \alpha_x/2,
\end{eqnarray}
where $g_x = {\rm diag} (e^{i\alpha_x/2},e^{-i\alpha_x/2})$. There exists a
$U(1)$ lattice gauge field, which is defined as
\begin{equation}
 A_{xy}  =  A^{\mu}_{x} \;
 = \,[-{\rm Arg} U_{xy}^{11} + \theta_{xy}]  \,{\rm mod} \,2\pi
\label{A}
\end{equation}
that transforms as $A_{xy}  \rightarrow  A_{xy} - \alpha_y + \alpha_x$. The
field $W$ transforms as $W_{xy}  \rightarrow  W_{xy}e^{-i\alpha_x}$.

In order to evaluate the masses of the $Z$-boson and the Higgs boson we use the
correlators:
\begin{equation}
\frac{1}{N^6} \sum_{\bar{x},\bar{y}} \langle \sum_{\mu} Z^{\mu}_{x} Z^{\mu}_{y}
\rangle   \sim
  e^{-M_{Z}|x_0-y_0|}+ e^{-M_{Z}(L - |x_0-y_0|)}
\label{corZ}
\end{equation}
and
\begin{equation}
  \frac{1}{N^6}\sum_{\bar{x},\bar{y}}(\langle H_{x} H_{y}\rangle - \langle H\rangle^2)
   \sim
  e^{-M_{H}|x_0-y_0|}+ e^{-M_{H}(L - |x_0-y_0|)},
\label{cor}
\end{equation}
Here the summation $\sum_{\bar{x},\bar{y}}$ is over the three ``space"
components of the four - vectors $x$ and $y$ while $x_0, y_0$ denote their
``time" components. $N$ is the lattice length in "space" direction. $L$ is the
lattice length in the "time" direction. In lattice calculations we used two
different operators that create Higgs bosons: $ H_x = |\Phi|$ and $H_x =
\sum_{y} Z^2_{xy}$. In both cases $H_x$ is defined at the site $x$, the sum
$\sum_y$ is over its neighboring sites $y$.

The physical scale is given in our lattice theory by the value of the $Z$-boson
mass $M^{phys}_Z \sim 91$ GeV. Therefore the lattice spacing is evaluated to be
$a \sim [91 {\rm GeV}]^{-1} M_Z$, where $M_Z$ is the $Z$ boson mass in lattice
units.  It has been found that the $W$ - boson mass contains an artificial
dependence on the lattice size. We suppose, that this dependence is due to the
photon cloud surrounding the $W$ - boson. The energy of this cloud is related
to the renormalization of the fine structure constant.
 Therefore the $Z$ - boson mass was used
in order to fix the scale.

Our data show that $\Lambda= \frac{\pi}{a} = (\pi \times 91~{\rm GeV})/M_Z$ is
increased slowly with the decrease of $\gamma$ at any fixed $\lambda$. We
investigated carefully the vicinity of the transition point at fixed $\lambda =
 0.009$ and $\beta = 12$. It has been found that at the transition point the
value of $\Lambda$ is equal to $1.4 \pm 0.2$ Tev. The check of the dependence
on the lattice size ($8^3\times 16$, $12^3\times 16$, $16^4$, $20^3\times 24$)
does not show an essential increase of this value (see Fig. \ref{fig.3}).

In the Higgs channel the situation is difficult. First, due to the lack of
statistics we cannot estimate the masses in this channel using the correlators
(\ref{cor}) at all considered values of $\gamma$. At the present moment at
$\lambda =0.009$ we can represent the data at the two points on the lattice
$8^3\times16$: ($\gamma = 0.274$, $\lambda =0.009$, $\beta = 12$) and ($\gamma
= 0.290$, $\lambda =0.009$, $\beta = 12$). The first point roughly corresponds
to the position of the transition while the second point is situated deep
within the Higgs phase.  At the point ($\gamma = 0.274$, $\lambda =0.009$,
$\beta = 12$) we have collected enough statistics to calculate correlator
(\ref{cor}) up to the "time" separation $|x_0-y_0| = 4$. The value $\gamma =
0.274$ corresponds roughly to the position of the phase transition. The mass
found in this channel in lattice units is $M^L_H = 0.75 \pm 0.1$ while bare
value of $M_H$ is $M^0_H \sim 270$ Gev. At the same time $M_Z^L = 0.23 \pm
0.007$. Thus we estimate at this point $M_H = 300 \pm 40$ Gev.  At the point
($\gamma = 0.29$, $\lambda =0.009$, $\beta = 12$) we calculate the correlator
with reasonable accuracy up to $|x_0-y_0| = 3$.  At this point bare value of
$M_H$ is $M^0_H \sim 260$ Gev while the renormalized Higgs mass in lattice
units is $M^L_H = 1.2 \pm 0.3$. At the same time $M_Z^L = 0.41 \pm 0.01$. Thus
we estimate at this point $M_H = 265 \pm 70$ Gev.

\section{Nambu monopole density }

The monopole density is defined as $ \rho = \left\langle \frac{\sum_{\rm
links}|j_{\rm link}|}{4V^L}
 \right\rangle,$
where $V^L$ is the lattice volume.
\begin{figure}
\begin{center}
 \epsfig{figure=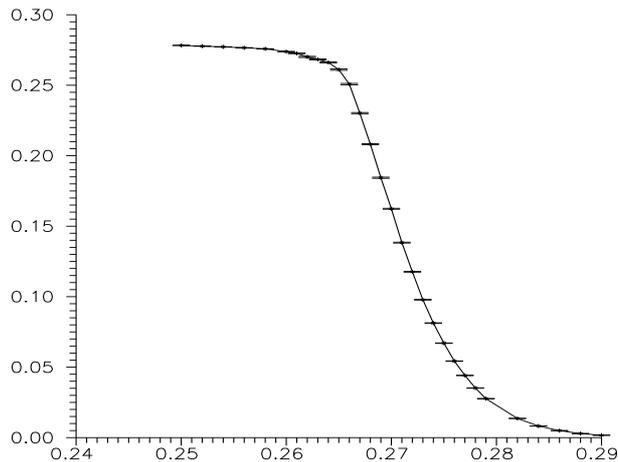,height=60mm,width=80mm,angle=0}
\caption{\label{fig.5_} Nambu monopole density as a function of $\gamma$ at
$\lambda =0.009$ and $\beta = 12$.  }
\end{center}
\end{figure}
On Fig \ref{fig.5_}  we represent Nambu monopole density as a function of
$\gamma$ at $\lambda = 0.009$, $\beta = 12$. The value of monopole density at
$\gamma_c$ is around $0.1$. At this point the value of the cutoff is $\Lambda
\sim 1.4 \pm 0.2$ Tev.

According to the classical picture the Nambu monopole size is of the order of
$M^{-1}_H$. Therefore, for example, for $a^{-1} \sim 430$ Gev and $M_H \sim
300, 150, 100$ Gev the expected size of the monopole is about a lattice
spacing. The monopole density around $0.1$ means that among $10$ sites there
exist $4$ sites that are occupied by the monopole. Average distance between the
two monopoles is, therefore, less than $1$ lattice spacing and it is not
possible at all to speak of the given configurations as of representing the
physical Nambu monopole. At $\gamma = \gamma_{c2}$ the Nambu monopole density
is around $0.03$. This means that among $7$ sites there exists one site that is
occupied by the monopole. Average distance between the two monopoles is,
therefore, approximately $2$ lattice spacings or $\sim \frac{1}{160\, {\rm
Gev}}$. Thus, the Nambu monopole density in physical units is around $[{160\,
{\rm GeV}}]^3$. We see that at this value of $\gamma$ the average distance
between Nambu monopoles is of the order of their size.

We summarize the above observations as follows. Within the fluctuational region
the configurations under consideration do not represent single Nambu monopoles.
Instead these configurations can be considered as the collection of monopole -
like objects that is so dense that the average distance between the objects is
of the order of their size. On the other hand, at $\gamma
>> \gamma_{c2}$ the considered configurations do represent single Nambu
monopoles and the average distance between them is much larger than their size.
In other words out of the FR vacuum can be treated as a gas of Nambu monopoles
while within the FR vacuum can be treated as a liquid composed of monopole -
like objects.

It is worth mentioning that somewhere inside the $Z$ string connecting the
classical Nambu monopoles the Higgs field is zero: $|\Phi| = 0$. This means
that the $Z$ string with the Nambu monopoles at its ends can be considered as
an embryo of the symmetric phase within the Higgs phase. We observe that the
density of these embryos is increased when the phase transition is approached.
Within the fluctuational region the two phases are mixed, which is related to
the large value of Nambu monopole density. That's why we come to the conclusion
that vacuum of lattice Weinberg - Salam model within the FR has nothing to do
with the continuum perturbation theory. This means that the usual perturbation
expansion around trivial vacuum (gauge field equal to zero, the scalar field
equal to $(\phi_m,0)^T$) cannot be valid within the FR.  This might explain why
we do not observe in our numerical simulations the large values of $\Lambda$
predicted by the conventional perturbation theory.

\section{Conclusions}

The continuum physics is to be approached in the vicinity of the phase
transition between the physical Higgs phase and the unphysical symmetric phase
of the model. The ultraviolet cutoff is increased when the transition point is
approached along the line of constant physics. There exists the so - called
fluctuational region (FR) on the phase diagram of the lattice Weinberg - Salam
model. This region is situated in the vicinity of the phase transition. We
calculate the effective constraint potential $V(\phi)$ for the Higgs field. It
has a minimum at the nonzero value $\phi_m$ in the physical Higgs phase. Within
the FR the fluctuations of the scalar field become of the order of $\phi_m$.
Moreover, the "barrier hight" $H = V(0) - V(\phi_m)$ is of the order of
$V(\phi_m + \delta \phi)- V(\phi_m)$, where $\delta \phi$ is the fluctuation of
$|\Phi|$.

The scalar field must be equal to zero somewhere within the classical Nambu
monopole. That's why this object can be considered as an embryo of the
unphysical symmetric phase within the physical Higgs phase of the model. We
investigate properties of the quantum Nambu monopoles. Within the FR they are
so dense that the average distance between them becomes of the order of their
size. This means that the two phases are mixed within the FR. All these results
show that the vacuum of lattice Weinberg - Salam model in the FR is essentially
different from the trivial vacuum used in the conventional perturbation theory.
As a result the use of the perturbation theory in this region is limited.

 Our numerical results show that at $M_H$ around $300$ GeV the maximal value of the cutoff admitted out of the FR
for the considered lattice sizes cannot exceed the value around $1.0 \pm 0.1$
Tev. Within the FR the larger values of the cutoff can be achieved. The
absolute maximum for the value of the cutoff within the Higgs phase of the
lattice model is achieved at the point of the phase transition. Our estimate
for this value is $1.4 \pm 0.2$ Tev for the lattice sizes up to $20^3\times
24$.

 {\bf Acknowledgment:}

This work was partly supported by RFBR grants 09-02-00338, 08-02-00661, by
Grant for leading scientific schools 679.2008.2. The numerical simulations have
been performed using the facilities of Moscow Joint Supercomputer Center.


\bigskip
\bigskip


\begin{thebibliography}{99}


\bibitem{BVZ2007} B.L.G. Bakker, A.I.
Veselov, M.A. Zubkov. J.Phys.G36:075008,2009.


\bibitem{Nambu}
Y.~Nambu, Nucl.Phys. B {\bf 130}, 505 (1977);\\
Ana~Achucarro and Tanmay~Vachaspati, Phys. Rept. {\bf 327}, 347 (2000); Phys.
Rept. {\bf 327}, 427 (2000).


\bibitem{BVZ2005}  B.L.G.
Bakker, A.I. Veselov, M.A. Zubkov. Phys.Lett.B620:156-163,2005.


\bibitem{BVZ2007_2}
B.L.G. Bakker, A.I. Veselov, M.A. Zubkov. PoS LAT2007:337,2007,.
[arXiv:0708.2864]

\bibitem{VZ2008}   A.I. Veselov, M.A.
Zubkov, JHEP 0812:109,2008.


\bibitem{Z2009}  M.A. Zubkov. Phys.Lett.B684:141-146,2010.

\bibitem{Chernodub_Nambu} M.N.~Chernodub, JETP Lett. {\bf 66}, 605 (1997)






\bibitem{Montvayold}
I.~Montvay, Nucl. Phys. B {\bf 269}, 170 (1986).

\bibitem{Montvay}
 W.Langguth, I.Montvay, P.Weisz
Nucl.Phys.B277:11,1986







\bibitem{forms}
M.I.~Polikarpov, U.J.~Wiese, and M.A.~Zubkov, Phys. Lett. B {\bf 309}, 133
(1993).









\bibitem{Kertesz}
M.N. Chernodub, Phys.Rev.Lett. 95 (2005) 252002


\end{thebibliography}
\end{document}